\documentclass[%
 reprint,
 superscriptaddress,
 amsmath,amssymb,
 aps,
 prb,
 longbibliography
]{revtex4-2}

\usepackage{graphicx}%
\usepackage{dcolumn}%
\usepackage{bm}%
\usepackage{color}%
\usepackage{siunitx}
\usepackage{here}
\usepackage{mathtools}
\usepackage{hyperref}
\newcommand{\red}[1]{\textcolor{black}{#1}}

\begin{document}
\preprint{APS/123-QED}

\title{Quantitative comparison of heat flow, guarded-heater and AC Harman methods for thermoelectric module efficiency}%

\author{K. Okawa}
\email{okawa.k@aist.go.jp}
\affiliation{%
National Metrology Institute of Japan (NMIJ), National Institute of Advanced Industrial Science and Technology (AIST), Tsukuba, Japan}%
\author{Y. Amagai}
\affiliation{%
National Metrology Institute of Japan (NMIJ), National Institute of Advanced Industrial Science and Technology (AIST), Tsukuba, Japan}%
\affiliation{%
Global Research and Development Center for Business by Quantum-AI Technology (G-QuAT), National Institute of Advanced Industrial Science and Technology (AIST), Tsukuba, Japan}%
\author{N. Sakamoto}
\affiliation{%
National Metrology Institute of Japan (NMIJ), National Institute of Advanced Industrial Science and Technology (AIST),  Tsukuba, Japan}%
\author{N-H. Kaneko}
\affiliation{%
National Metrology Institute of Japan (NMIJ), National Institute of Advanced Industrial Science and Technology (AIST), Tsukuba, Japan}%
\affiliation{%
Global Research and Development Center for Business by Quantum-AI Technology (G-QuAT), National Institute of Advanced Industrial Science and Technology (AIST), Tsukuba, Japan}%
\

\begin{abstract}
The evaluation of thermoelectric conversion efficiency remains challenging owing to the lack of internationally standardized measurement protocols. Commonly used techniques—including the heat flow, guarded heater, and AC Harman methods—differ fundamentally in their operating principles and sensitivity to heat losses. In this study, we benchmark three module-level efficiency measurement techniques—the heat-flow, guarded heater, and AC Harman methods—using commercial $\rm Bi_{2}Te_{3}$-based modules \red{with different module architectures. The conversion efficiencies obtained using the heat flow and guarded heater methods showed closely consistent central values and similar temperature-dependent trends over the investigated range. In contrast, the efficiency derived using the AC Harman method was systematically lower by approximately \SI{16}{\%} to \SI{30}{\%}, depending on the module architecture. Steady-state finite-element calculations of heat conduction and radiation indicated that the open thermal boundary condition used in the Harman configuration produces module-architecture-dependent internal temperature distributions and effective temperature differences, consistent with the experimentally observed trend. These results demonstrate that module-level efficiency estimated using the AC Harman method can be affected by nonideal thermal environments and emphasize the necessity of accounting for radiative and substrate-related heat losses. Nevertheless, the AC Harman method remains useful for rapid performance screening, provided that its module- and boundary-condition-dependent systematic bias is appropriately considered.} Our results provide a quantitative benchmark for major measurement techniques and support the development of best practices, method-selection guidelines, and future methodological standardization in module-level thermoelectric metrology.
 \end{abstract}

\maketitle
\section{Introduction}
\label{introduction}
Thermoelectric modules play a key role in waste heat recovery by efficiently converting waste heat into electrical energy. Dissemination of thermoelectric power generation using waste heat necessitates the establishment of standardized and accurate measurement techniques to precisely estimate conversion efficiency \cite{Heremans2024,Bell2008,Wei2018,Carducci2020,Rowe1995}. The uncertainty in measuring thermoelectric properties is approximately $\SI{20}{\%}$ as per the results of a recent international round-robin test \cite{Wang2013,Wang2015,Alleno2015}; the results highlight the lack of a standardized evaluation protocol or sample modules in the field of thermoelectrics\cite{Wang2014,Ziolkowski2022,Ziolkowski2020,Ziolkowski2021BiTe}.

\red{The performance of thermoelectric power generation modules is typically evaluated by assessing their conversion efficiency, $\eta$, defined as the ratio of the electrical output power, $P_{\rm out}$, to the input heat flow to the hot-side of the module, $Q_{\rm in}$:}
\begin{eqnarray}
\eta = \frac{P_{\rm out}}{Q_{\rm in}}.
\label{eq1}
\end{eqnarray}
Accurate evaluation of $Q_{\rm in}$ is crucial for determining conversion efficiency and is a major challenge\cite{Ziolkowski2021}. Two typical methods for evaluating the conversion efficiency of thermoelectric modules are the heat flow method (HFM) and guarded heater method (GHM), which differs in their approaches to assessing $Q_{\rm in}$\cite{Wang2014,Raju2020,Ziolkowski2021}.

The AC Harman method, which estimates the dimensionless figure of merit, $ZT$, through dc and ac resistance measurements, provides a rapid evaluation of module-level performance\cite{Harman1958,Iwasaki2003}. The heat flow and guarded heater methods are steady-state techniques, whereas the AC Harman method utilizes the transient thermal response of the module. \red{Under DC excitation, the Peltier effect establishes a temperature difference across the thermoelectric element, and the resulting Seebeck voltage contributes to the measured voltage. The measured resistance therefore approaches $R_{\rm{DC}}$, which contains both the ohmic and thermoelectric contributions. At a sufficiently high ac frequency, the Peltier-induced temperature oscillation is suppressed, and the measured resistance approaches the purely ohmic resistance, $R_{\rm {AC}} \simeq R_{\rm {ohm}}$. Under the ideal Harman assumptions, the effective module figure of merit is evaluated as}
\begin{eqnarray}
ZT = \frac{R_{\rm DC}-R_{\rm AC}}{R_{\rm AC}}.
\label{eq5}
\end{eqnarray}
\red{Here, $R_{\mathrm{AC}}$ represents the high-frequency limiting resistance for which the thermoelectric contribution is sufficiently suppressed, rather than the resistance measured at an arbitrary AC frequency.}
\begin{figure*}[t]
\centering
\includegraphics[width=17.8cm,clip]{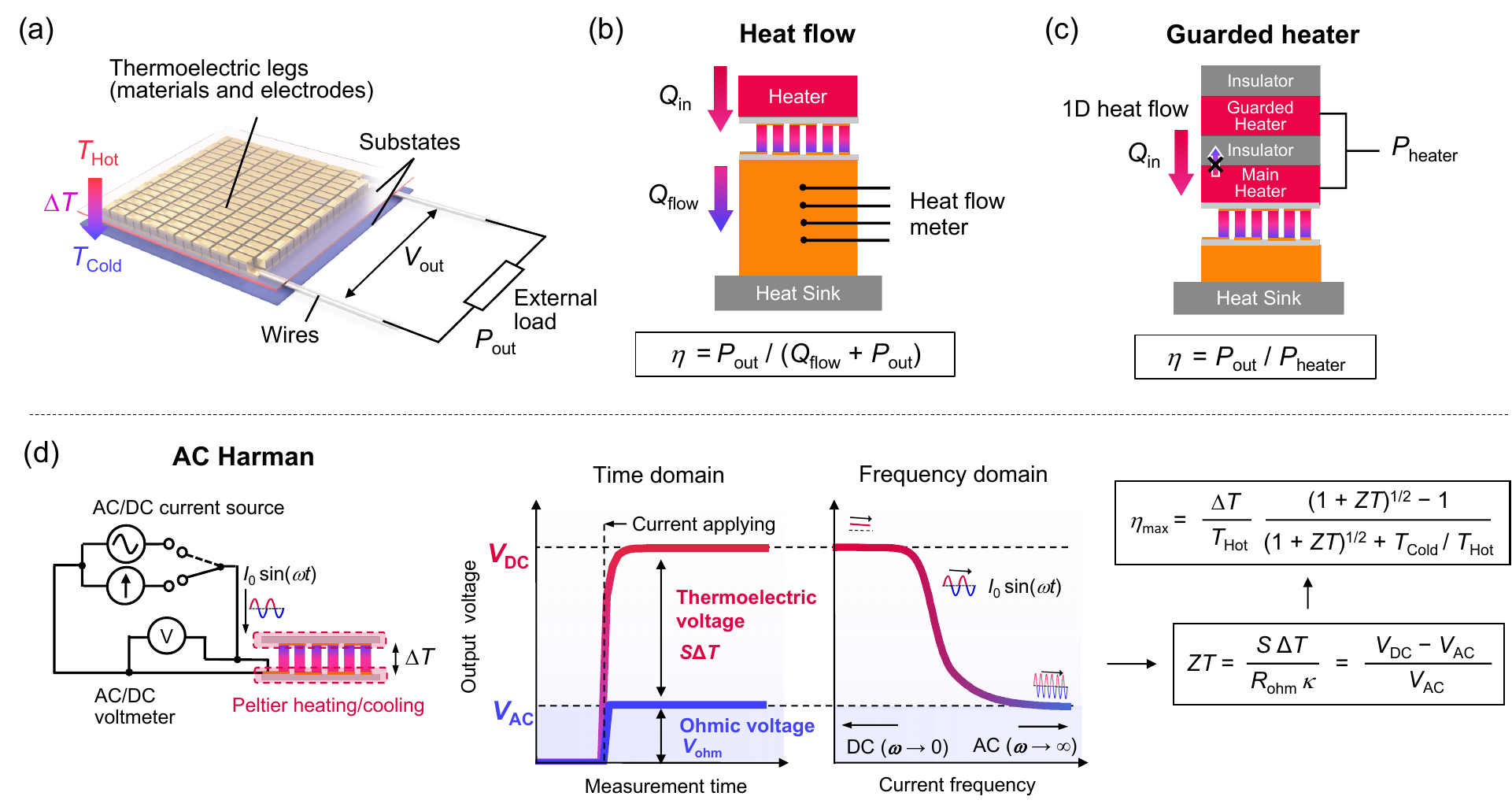}
\caption{\red{(a) Schematic of a typical thermoelectric module. A temperature difference, $\Delta T = T_{\rm Hot}-T_{\rm Cold}$, is maintained between the hot and cold surfaces. The thermoelectric legs are electrically connected in series and thermally connected in parallel between substrates. The resulting Seebeck voltage drives an electrical current through an external load, delivering an output power, $P_{\rm out}$. (b-d) Schematics of the three representative techniques for evaluating thermoelectric conversion efficiency. (b) In the heat flow method, the heat passing through the module, $Q_{\rm flow}$, is measured using a heat-flow meter, and the efficiency is determined as $\eta = P_{\rm out}/(Q_{\rm flow}+P_{\rm out})$. (c) In the guarded heater method, parasitic lateral heat losses are suppressed by the guard heater so that approximately one-dimensional heat flow is established; the efficiency is evaluated from the heater power as $\eta = P_{\rm out}/P_{\rm heater}$. (d) In the AC Harman method, the voltage responses to DC and sufficiently high-frequency AC currents are used to separate the thermoelectric contribution from the ohmic response. The DC voltage, $V_{\rm DC}$, contains both the ohmic voltage and the thermoelectric voltage generated by the Peltier-induced temperature difference. At a sufficiently high AC frequency, the thermal response is suppressed, and the measured voltage, $V_{\rm AC}$, represents the predominantly ohmic response. The effective module figure of merit is obtained from $ZT = (V_{\rm DC} - V_{\rm AC})/V_{\rm AC}$, and the corresponding maximum conversion efficiency is subsequently calculated using $ZT$ and the hot- and cold-side temperatures.}}
\label{fig1}
\end{figure*}
This method is widely used in estimating $ZT$ for various samples, including minute crystals, thin films, and composite structures, owing to its simplicity and reliance on a simple two-probe configuration. \red{At the module level, the measured response can be influenced by substrates, metallic interconnects, electrical and thermal contacts, lead wires, module geometry, and heat exchange with the surroundings. The resulting value should therefore be regarded as an effective module-level figure of merit under the specific measurement boundary conditions, rather than as an intrinsic in-plane material property or a direct measurement of conversion efficiency.} 

\red{Based on the obtained effective $ZT$, the corresponding maximum conversion efficiency, $\eta_{\rm max}$, is estimated using the conventional constant-property expression\cite{Rowe1995}:}
\begin{eqnarray}
\eta_{\rm max} = \frac{\Delta T}{T_{\rm Hot}}\frac{\sqrt{(1+ZT)}-1}{\sqrt{(1+ZT)}+T_{\rm Cold}/T_{\rm Hot}},
\label{eq2}
\end{eqnarray}
where $T_{\rm Hot}$ and $T_{\rm Cold}$ are the hot- and cold-side temperatures, respectively.

Comparison with direct measurement techniques such as the HFM and GHM is essential for identifying potential sources of bias and improving measurement accuracy. \red{However, a systematic quantitative comparison of the HFM, GHM, and AC Harman methods under comparable module-level conditions has not yet been reported\cite{Rauscher2003,Takazawa2006,Pierce2015,Zhu2022}. A recent study instead focused on material-level characterization\cite{He2021}.} In particular, the extent to which differences in measurement principles and thermal boundary conditions lead to discrepancies in evaluated conversion efficiency remains unclear.

Here, we perform a comparison of the three measurement techniques for characterizing the conversion efficiency using reliable commercial $\rm Bi_{2}Te_{3}$-based modules with different types of substrates. To perform the three types of measurements, we have developed an apparatus capable of evaluating these methods. \red{The discrepancies observed in the evaluated conversion efficiencies obtained from the three measurement techniques are hypothesized to originate from differences in the boundary conditions of the measurement environment and the actual temperature gradients established within the module.} To investigate these factors, a combined analysis of experimental results and finite element method (FEM) simulations was performed using COMSOL Multiphysics\cite{COMSOL}.

\section{Experimental setup}
\subsection{Sample information}
The commercial thermoelectric modules exhibit \red{variations in area, height, thermoelectric-leg geometry, number of p-n pairs, and substrate configuration} as shown in Fig.~1(a). The properties of the modules used in this study are presented in Table~1. This evaluation employed commercially available Bi-Te modules, including a skeleton type with a polyimide substrate measuring $\SI{48}{mm} \times \SI{56.6}{mm} \times \SI{1.36}{mm}$, and a rigid type with an alumina substrate measuring $\SI{51.2}{mm} \times \SI{55.0}{mm} \times \SI{4.4}{mm}$. The number of p-n pairs in each thermoelectric module is 199 for the skeleton type and 161 for the rigid type. The electrode thicknesses are $\SI{0.06}{mm}$ for the skeleton-type module and $\SI{0.60}{mm}$ for the rigid alumina-substrate module.
\begin{table*}[]
\centering
\begin{tabular}{l c c}
Parameter & Skeleton-type module & Rigid alumina-substrate module \\ 
\hline
  Substrate material & Polyimide & Alumina \\
  Module dimensions $(\rm mm^{3})$ & $48.0 \times 56.6 \times 1.36$ & $51.2 \times 55.0 \times 4.4$ \\
  Number of p-n pairs & 199 & 161 \\
  Electrode thickness (mm) & 0.06 & 0.60 \\
  \hline
\end{tabular}
\caption{Properties of thermoelectric modules}\label{Table1}
\end{table*}
\subsection{Heat flow and guarded heater methods}
In HFM, a calibrated Cu-block heat flow meter was employed to measure heat flow from the cold side of the module $Q_{\rm flow}$\cite{Ziolkowski2021,He2021}. Embedded thermocouples along the Cu block accurately measured the temperature gradient. The heat flow was calculated using the Fourier equation. The input heat flow, $Q_{\rm in}$, into the module can be determined along with the maximum output power, $P_{\rm out}$, generated by the module using an external electrical load device as: $Q_{\rm in} = Q_{\rm flow} + P_{\rm out}$. The combined standard uncertainty of heat flow evaluation in the HFM ranges from \SI{10}{\%} to \SI{13}{\%} (\SI{373}{K} to \SI{1023}{K}) at a coverage factor $k = 2$, corresponding to a confidence level of \SI{95}{\%}\cite{Ziolkowski2021}. As shown in Fig. 1(b), the HFM involves placing a thermoelectric module in series with a reference block of known thermal conductivity. Multiple thermometers are embedded in the reference block, and heat flow is evaluated based on the temperature gradient.

\red{Details of the basic apparatus design and its application to HFM/GHM measurements were reported previously\cite{Okawa2020}. The present work used the same basic measurement principle and apparatus configuration, with the temperature range and sample modules adapted to the present comparison.} The apparatus utilized HFM and GHM to accurately measure the output power and heat flow in a single system. A schematic of the apparatus is shown in Fig. 2(a). The temperature of the hot side of the module was varied in the range of \SI{303}{K} to \SI{413}{K} using heaters, whereas the cold side temperature was maintained at \SI{303}{K} with a water-cooling heat sink. Thermocouples embedded in the Cu block were used to measure the actual temperatures at both ends. Therefore, the thermoelectric conversion efficiency $\eta$ can be expressed as follows:
\begin{eqnarray}
\eta = \frac{P_{\rm out}}{Q_{\rm flow}+P_{\rm out}}.
\label{eq3}
\end{eqnarray}
GHM determines the input heat flow, $Q_{\rm in}$, based on the electric power input to the heater, $P_{\rm heater}$\cite{Rauscher2003,Rauscher2005}. \red{The GHM employs two heaters with different functions.  This principle is based on the construction of a stack consisting of a polymer insulating block and a guard heater, which directs the input heat current in a single direction (Fig. 1(c)). A guard heater set at the same temperature as the main heater is positioned above the main heater, which is located on top of the thermoelectric module with an insulating plate in between. The guard heater is controlled at approximately the same temperature to suppress lateral heat flow from the main heater.}  Assuming that all the input heat conducts through the module, $Q_{\rm in}$ can be estimated as $P_{\rm heater}$ ($Q_{\rm in} = P_{\rm heater}$). Therefore, $\eta$ can be expressed as follows:
\begin{eqnarray}
\eta = \frac{P_{\rm out}}{P_{\rm heater}}.
\label{eq4}
\end{eqnarray}
In principle, the combined standard uncertainty of the evaluation of the input heat flow Qin for GHM is estimated to be approximately \SI{0.1}{\%} to \SI{0.8}{\%} ($k = 2$)\cite{Ziolkowski2021}. 

\red{The HFM and GHM directly determine conversion efficiency and do not directly measure $ZT$. For comparison with the AC Harman result, the HFM and GHM efficiency data were fitted using Eq. (3), assuming a temperature independent effective $ZT$ over the investigated range. The resulting fitted values are efficiency equivalent module-level parameters and should not be interpreted as intrinsic local material $zT$($T$)}.

The heater was powered by a DC power supply (Kikusui PAN110-10A) under PID control to maintain the target temperature. The $Q_{\rm flow}$ was measured using thermocouples embedded in a Cu block, with the signals simultaneously recorded by a Keithley 3706A system switch/multimeter equipped with a 3721 dual $1 \times 20$ multiplexer card. Measurements of power generation properties were performed in vacuum using an electronic load system (Kikusui PLZ164WA), in which current-voltage ($I$-$V$) and current-power ($I$-$P$) characteristics were obtained by sweeping the load resistance. In both methods, substantial attention was given to minimizing errors by ensuring good thermal contact between the module and Cu blocks, as well as reducing heat loss from the heater block and heat sink. Thermal grease was applied between the Cu block and module to optimize thermal contact. \red{To evaluate the reproducibility of the thermal contact, the dependence of the module output voltage on the applied compressive force was examined. Above approximately \SI{1000}{N}, the output voltage became independent of the compressive force. This observation supports the reproducibility of the mounting condition above approximately \SI{1000}{N}.}

\noindent

\begin{figure}[t]
\centering
\includegraphics[width=8cm,clip]{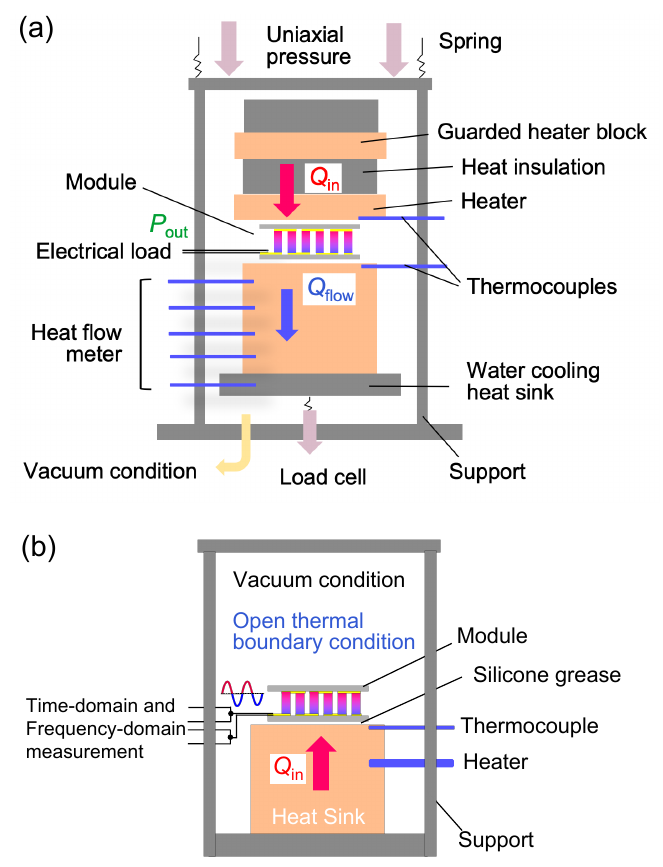}
\caption{\red{Schematic illustrations of the measurement apparatus. 
(a) Experimental setup used for the heat flow method (HFM) and guarded heater method (GHM), in which the module is compressed between the heater block and the water-cooled heat sink. 
(b) Experimental setup used for the AC Harman method under vacuum conditions, in which the module is mounted on a heater block with an open thermal boundary condition at the opposite side.}}
\label{fig2}
\end{figure}

\subsection{AC Harman method}
The AC Harman method enables rapid estimation of the effective module-level $ZT$ from the electrical response of the module (Fig.~1(d))\cite{Harman1958,Iwasaki2003}. \red{The measured voltage consists of an ohmic voltage and a thermoelectric voltage generated by the Peltier-induced temperature difference. Under DC excitation, or equivalently in the sufficiently low-frequency limit in which the thermal response is fully developed, the measured resistance can be expressed as \begin{equation} R_{\mathrm{DC}} = R_{\mathrm{ohm}} + \frac{S\Delta T_{\mathrm{Peltier}}}{I}, \end{equation} where $R_{\mathrm{ohm}}$ is the purely ohmic resistance, $S$ is the Seebeck coefficient, $I$ is the applied current, and $\Delta T_{\mathrm{Peltier}}$ is the temperature difference generated by Peltier heating and cooling. As the current frequency increases, the Peltier-induced temperature response is progressively suppressed because thermal diffusion can no longer follow the current modulation \cite{Okawa2024,Yoo2019,Hasegawa2022,Hasegawa2023}. At a sufficiently high frequency, the thermoelectric contribution to the measured voltage becomes negligible, and the measured AC resistance approaches the ohmic resistance: 
\begin{equation} R_{\mathrm{AC}} \simeq R_{\mathrm{ohm}}. \end{equation} 
Here, $R_{\mathrm{AC}}$ denotes the high-frequency limiting resistance and not the resistance measured at an arbitrary AC frequency. The AC frequency was considered sufficiently high when the real part of the measured impedance reached a high-frequency resistance plateau, indicating that the Peltier-induced thermal response was sufficiently suppressed. The appropriate frequency is sample- and system-dependent and should be selected such that the Peltier-induced response is suppressed while parasitic electrical dispersion remains negligible. }

\red{Following the conventional Harman approximation, the purely ohmic resistance is assumed to be effectively independent of frequency between the DC and high-frequency limiting regions. Under these assumptions, the effective module figure of merit was evaluated from the difference between the DC and high-frequency AC limiting resistances using Eq.~(2). Both time-domain and frequency-domain measurements were performed to verify the limiting resistance values. In the time-domain measurement, the DC voltage included both the ohmic and thermoelectric contributions, whereas the voltage measured using sufficiently high-frequency AC excitation represented the predominantly ohmic response. In the frequency-domain measurement, $R_{\mathrm{DC}}$ and $R_{\mathrm{AC}}$ were identified from the low- and high-frequency limiting regions, respectively, of the real part of the measured impedance.}

The experimental setup used for the AC Harman method, as shown in Fig. 2(b), employed a heater block placed beneath the module to control its temperature. \red{In the AC Harman configuration, the module was mounted on a temperature controlled heater block, whereas the opposite surface was not connected to a temperature controlled heat sink. The temperature of the exposed side was therefore determined by the balance among internal heat conduction, radiation, and any remaining parasitic heat-transfer paths. This open thermal boundary differs fundamentally from the prescribed hot- and cold-side temperatures used in the HFM and GHM configurations. The current amplitude was selected to obtain an adequate thermoelectric signal while suppressing the contribution of Joule heating.}
Silicone grease was applied between the module and the heater block to ensure good thermal contact. 

In previous studies, we evaluated BiTe-based thermoelectric materials using a comparable approach and found that the $(R_{\rm DC}-R_{\rm AC})/R_{\rm AC}$ value measured at ambient pressure was underestimated by up to \SI{40}{\%} compared with measurements performed under high-vacuum conditions, owing to increased convective heat loss\cite{Okawa2024}. \red{The underestimation at ambient pressure was attributed to greater gas-mediated heat transfer, which reduced the Peltier-induced temperature difference compared with that under high-vacuum conditions.} In the present apparatus, when operated under vacuum conditions with pressures $P \leq \SI{e-1}{Pa}$, \red{gas-mediated heat transfer was suppressed, although the remaining residual-gas contribution was not independently quantified.} 

For the time-domain measurements, DC and AC currents with identical root-mean-square values were applied to the module using an AC/DC current source (Keithley 6221), and the resulting voltages were measured using a nanovoltmeter (Keithley 2182A). The DC voltage was used to determine the response containing both the ohmic and thermoelectric contributions, whereas the voltage measured at the selected AC frequency was used to determine the predominantly ohmic response. \red{For the frequency-domain measurements, a HIOKI IM3590 impedance analyzer was used to measure the frequency dependence of the complex impedance. The integration time was set to one cycle of the applied signal. The low- and high-frequency limiting resistances were identified from the corresponding limiting regions of the real part of the measured impedance. The selected value of $R_{\mathrm{AC}}$ was taken from the high-frequency resistance region in which the thermoelectric contribution was suppressed and no appreciable additional electrical dispersion was observed.}

\red{Each AC Harman measurement was repeated five times under the same current, frequency, temperature, and vacuum conditions. The reported values are the arithmetic means of the five measurements, and the repeatability was evaluated using the sample standard deviation.}
\noindent

\section{Numerical simulation}
The temperature distributions of the thermoelectric modules were calculated by finite-element analysis using COMSOL Multiphysics$^{\circledR}$\cite{Buchalik2024,Yang2024}.  The Heat Transfer in Solids interface was coupled with the Surface-to-Surface Radiation interface.
The governing equation for steady-state heat conduction in the solid domain is:
\begin{eqnarray}
\nabla \cdot (-\kappa(T)\nabla T) = Q
\end{eqnarray}
where $\kappa$($T$) is the temperature-dependent thermal conductivity obtained from the COMSOL materials database, $T$ is the temperature, and $Q$ represents internal volumetric heat sources, if any. Surface-to-Surface radiation between diffuse and gray surfaces was modeled using the radiosity method. The net radiative heat flux $q_{\rm rad}$ at a surface is defined as:
\begin{eqnarray}
q_{\rm rad} = \epsilon(J-G)
\end{eqnarray}
where $\epsilon$ is the surface emissivity, $J$ is the radiosity, and $G$ is the incident irradiation. These are given by:
\begin{eqnarray}
J &=& \epsilon \sigma T^{4} + (1-\epsilon)G, \\
G &=& \sum_{j = 1}^{N}F_{ij}J_{j}.
\end{eqnarray}
Here, $\sigma$ is the Stefan-Boltzmann constant and $F_{ij}$ is the view factor between surfaces $i$ and $j$\cite{Barry2016}. The computed net radiative heat fluxes were incorporated as boundary heat sources in the conduction equation, achieving full coupling between conduction and radiation. The thermal radiation with a module includes the radiation between the hot- and cold-junction substrates and the radiation among the side surfaces of thermoelectric legs and the exposed surfaces of electrodes\cite{Bjork2014}. Both surface-to-surface radiation and surface-to environment radiation must be considered. Furthermore, the position relations among all radiation surfaces and the interference by other surfaces are complicated. To accurately estimate the radiation heat losses existing in a module, the analytical physical equations are difficult to construct and solve\cite{Cai2020,Qing2024}. The model geometry was three-dimensional, and the computational domain was discretized using a tetrahedral mesh generated by COMSOL’s physics-controlled meshing algorithm. Local mesh refinement was applied in regions exhibiting large thermal gradients or complex radiative interactions. Convergence was confirmed by monitoring residual norms and verifying mesh independence. 
\begin{figure}[t]
\centering
\includegraphics[width=7cm,clip]{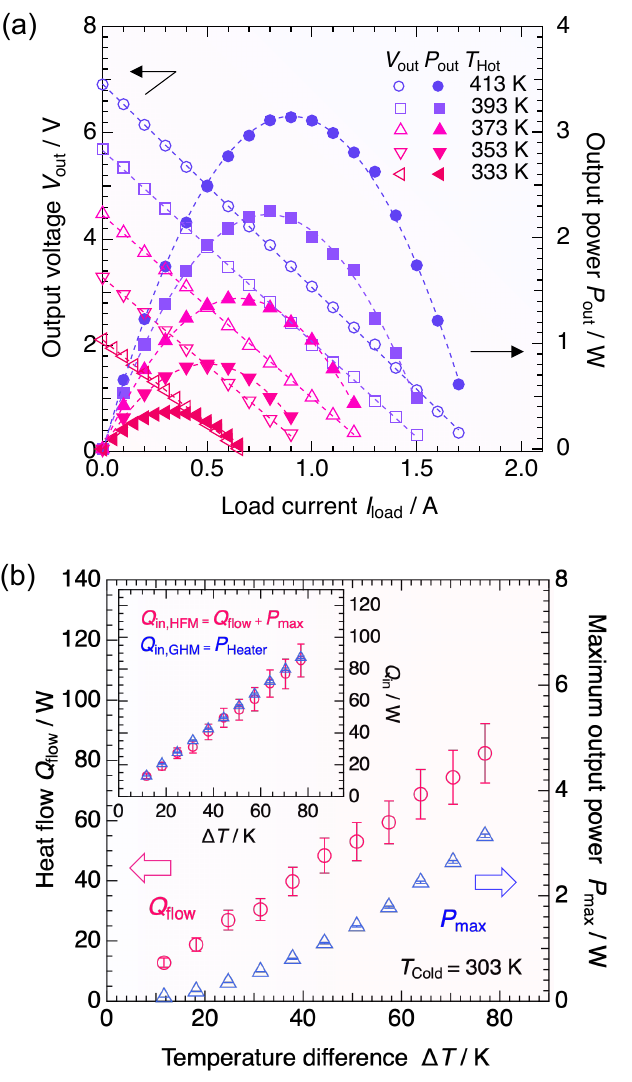}
\caption{\red{Power-generation characteristics and heat-flow evaluation of the thermoelectric module. (a) Output voltage, $V_{\rm out}$, and output power, $P_{\rm out}$, as functions of load current, $I_{\rm load}$, at different hot-side temperatures.  (b) Heat flow from the cold side of the module, $Q_{\rm flow}$, and maximum output power, $P_{\rm max}$, as functions of temperature difference, $\Delta T$, with the cold-side temperature maintained at $T_{\rm Cold} = \SI{303}{K}$. The inset compares the input heat flow evaluated using the heat flow method, $Q_{\mathrm{in,HFM}} = Q_{\rm flow} + P_{\rm max}$, and the guarded heater method, $Q_{\mathrm{in,GHM}} = P_{\rm Heater}$.}}
\label{fig3}
\end{figure}

\red{The present model was limited to steady-state heat conduction and radiative heat exchange under prescribed thermal boundary conditions. Current-induced Peltier heating, Joule heating, electrical transport, and the transient response of the AC Harman measurement were not explicitly included. Consequently, the FEM results were used to interpret relative differences in the internal temperature distributions of the two module configurations, rather than to calculate a direct correction factor for the measured Harman $ZT$.}

\red{Ideal thermal contact was assumed at the interfaces between the module and the heater block and between the internal model components. Because the thermal contact resistance was not independently measured, this assumption introduces an additional uncertainty in the calculated absolute interface temperatures. The FEM results are therefore interpreted primarily as a relative comparison between the module configurations under identical contact assumptions.}

\section{Results and discussion}
\subsection{Experimental results}
The power-generation characteristics of the skeleton-type thermoelectric module were evaluated using the experimental setup shown in Fig.~2(a). Figure~3(a) shows the measured current-voltage characteristics and the corresponding output power at various temperature differences, $\Delta T$, across the module. The open-circuit voltage of the skeleton-type module, $V_{\rm out}$, reached \SI{6.2}{V}, and the maximum output power, $P_{\rm max}$, reached \SI{3.2}{W} at $\Delta T = \SI{80}{K}$. These measured values were consistent with the catalog specifications. The maximum output power, $P_{\rm max}$, exhibited a quadratic dependence on $\Delta T$ ($\propto \Delta T^2$), as shown in Fig. 3(b).
\red{The measured input heat flow as a function of $\Delta T$, obtained using the HFM and the GHM, is also shown in Fig. 3(b). The inset of Fig. 3(b) compares the input heat flow evaluated by the two methods, $Q_{\rm in,HFM} = Q_{\rm flow} + P_{\rm max}$ and $Q_{\rm in,GHM} = P_{\rm Heater}$. The two estimates showed similar temperature dependence and agreed closely over the investigated temperature range, supporting the consistency between the HFM and GHM evaluations. Having evaluated the thermoelectric module’s performance, we next investigated the thermoelectric conversion efficiency using the setup shown in Fig. 2(b).}

The results of the AC Harman method are shown in Fig. 4. Figure 4(a) shows the time-dependent voltage responses under DC and AC excitation. The measurements were performed with the heater temperature set to \SI{373}{K}, an applied current of \SI{10}{mA}, and an AC frequency of \SI{10}{Hz}. The measured dc and ac voltages were \SI{64}{mV} and \SI{36}{mV}, respectively. The DC voltage contained both the ohmic and Peltier-induced thermoelectric contributions, whereas the response at \SI{10}{Hz} was predominantly ohmic because the thermal response was sufficiently suppressed at this frequency. Substitution of these values into Eq. (2) yielded an effective module $ZT$ of 0.81.

\red{Figure 4(b) shows the frequency dependence of the real part of the measured impedance. The measurement was performed at a heater temperature of \SI{373}{K} and an applied current of \SI{10}{mA}. At frequencies below \SI{1}{mHz}, the resistance approached the DC limiting value, $R_{\mathrm{DC}}$, indicating that the Peltier-induced thermal response was fully developed. With increasing frequency, the thermoelectric contribution decreased, and the measured resistance approached the high-frequency AC limiting value, $R_{\mathrm{AC}}$, where the response was predominantly ohmic. The use of this high-frequency resistance region is consistent with the approximation $R_{\mathrm{AC}} \simeq R_{\mathrm{ohm}}$.}

\red{Substitution of the independently determined $R_{\mathrm{DC}}$ and $R_{\mathrm{AC}}$ values into Eq. (2) yielded an effective module $ZT$ of 0.81, in agreement with the time-domain result. This agreement supports the selection of the \SI{10}{Hz} excitation used for the time-domain measurement and the identification of the corresponding DC and high-frequency limiting resistance regions.} The corresponding Nyquist plot is shown in the inset of Fig. 4(b).

The frequency response of the thermoelectric module was obtained using impedance spectroscopy\cite{Yoo2019}. Because Joule heating is time dependent, the resistance at each frequency was evaluated over the duration of one complete measurement cycle. \red{In general, the current amplitude should be selected by balancing the signal-to-noise ratio against the increase in Joule heating\cite{Okawa2024,Hasegawa2022,Hasegawa2023}. In the present study, the current was fixed at \SI{10}{mA} rms.}

Using the obtained $ZT$ values, the temperature dependence of the module conversion efficiency was estimated using Eq. (3). In this analysis, the temperature differential between the hot and cold sides, $\Delta T$ was assumed to be identical to that obtained from the other two methods. \red{Unlike the HFM and GHM, the AC Harman method did not directly measure $Q_{\rm in}$ or conversion efficiency. The plotted efficiency was calculated from the measured effective $ZT$ using Eq. (3), with the prescribed hot- and cold-side temperatures.} 

Figure 5 shows the temperature-difference dependence of the conversion efficiency of the skeleton-type module evaluated using the three different methods. Representative literature-based uncertainty values were adopted for the HFM and GHM results based on Ziolkowski et al.~\cite{Ziolkowski2022}. Accordingly, the error bars were taken as \SI{12}{\%} for the HFM and \SI{1}{\%} for the GHM. The dominant contribution to the adopted HFM uncertainty arises from heat-flow evaluation.The HFM and GHM results exhibited similar temperature dependence and differed by less than approximately \SI{2}{\%} up to a temperature difference of \SI{64}{K}. Both datasets also lay within the literature-based reference uncertainty ranges adopted for the two methods. \red{For the AC Harman method, each measurement was repeated five times, and the error bars represent one sample standard deviation of the repeated measurements. Because these uncertainty representations have different definitions, they are not directly comparable as equivalent confidence intervals and are therefore not used here to establish formal statistical agreement among the three methods. Instead, the measured central values and their temperature-dependent trends are compared quantitatively.}

\red{The AC Harman repeatability was small compared with the differences among the measurement methods. The repeatability estimate does not include systematic bias associated with nonideal thermal boundary conditions. The relative standard deviation was \SI{0.02}{\%}, substantially smaller than the difference between the AC Harman-derived efficiency and the directly measured efficiencies. In several cases, the AC Harman error bars are smaller than the symbol size.}

\begin{figure}[t]
\centering
\includegraphics[width=7cm,clip]{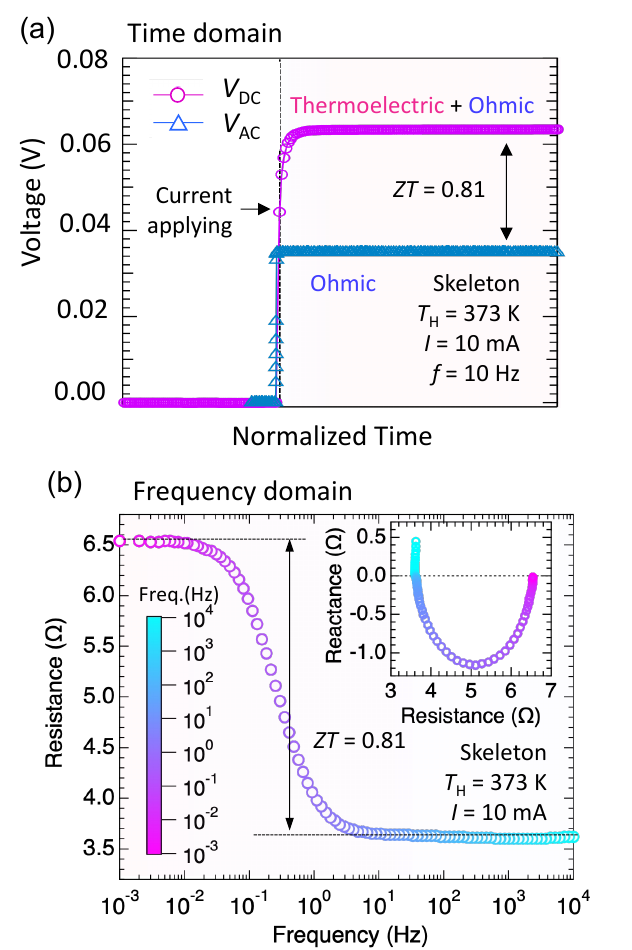}
\caption{Experimental results for the skeleton-type module. (a) Time-dependent dc and ac voltage responses. The heater temperature was $T_{\rm H} = \SI{373}{K}$, the applied current was $I = \SI{10}{mA}$, and the AC frequency was $f = \SI{10}{Hz}$. The DC response contains both the ohmic and thermoelectric contributions, whereas the \SI{10}{Hz} AC response is predominantly ohmic. \red{(b) Frequency dependence of the real part of the measured impedance under $T_{\rm H} = \SI{373}{K}$ and $I = \SI{10}{mA}$. The low-frequency limit corresponds to $R_{\mathrm{DC}}$, for which the Peltier-induced thermal response is fully developed, and the high-frequency limiting region corresponds to $R_{\mathrm{AC}}\simeq R_{\mathrm{ohm}}$.} The inset shows the corresponding Nyquist plot.}
\label{fig4}
\end{figure}
\begin{figure}[t]
\centering
\includegraphics[width=7cm,clip]{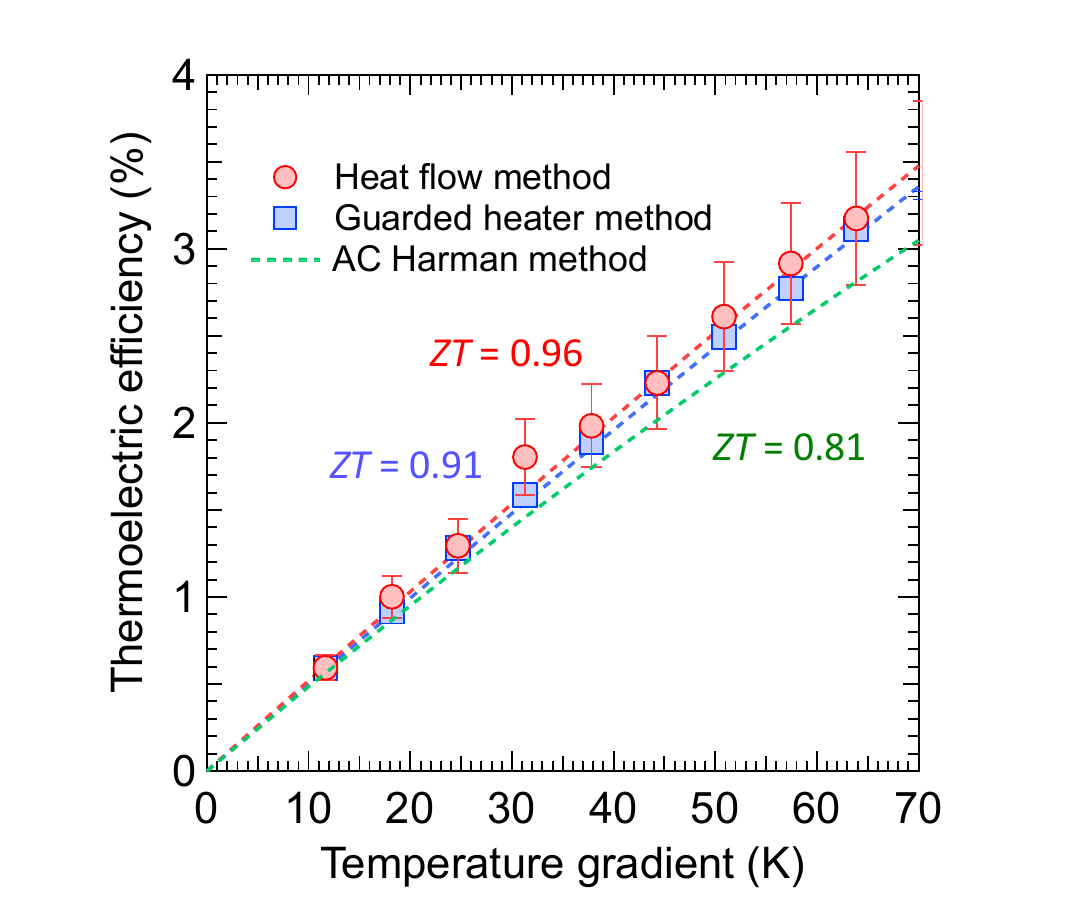}
\caption{Temperature-difference dependence of the thermoelectric conversion efficiencies of the skeleton-type module using the heat flow method (HFM), guarded heater method (GHM), and AC Harman method. Red circles and blue squares represent the values obtained using the heat flow and guarded heater methods, respectively. The dotted lines indicate the conversion efficiency curves predicted by the model based on Eq. (3). The corresponding $ZT$ values were obtained from the fitting results using Eq. (3). \red{The HFM and GHM error bars indicate representative literature-based expanded uncertainty ranges adopted in this study, whereas the AC Harman error bars represent one sample standard deviation of five repeated measurements. Because these error bars have different definitions, they should not be interpreted as directly comparable confidence intervals. The AC Harman error bars are smaller than the symbol size (less than \SI{0.02}{\%}). The AC Harman values are efficiencies estimated from the measured effective $ZT$, whereas the HFM and GHM values were obtained from direct measurements of electrical power and heat input.}}
\label{fig5}
\end{figure}

\begin{figure*}[t]
\centering
\includegraphics[width=16.5cm,clip]{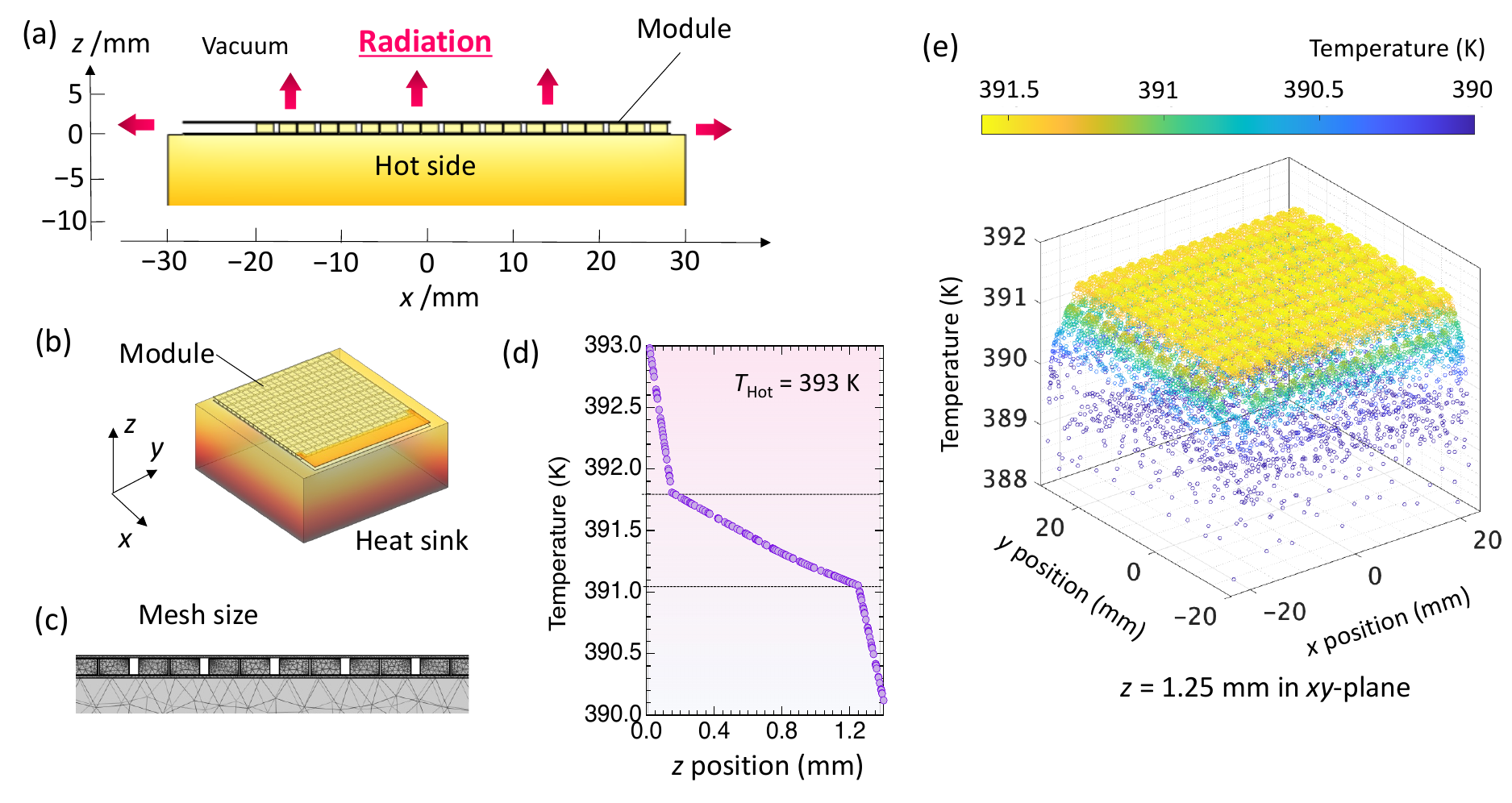}
\caption{Results of numerical simulations performed using the finite element method. (a) Schematic cross-sectional view of the skeleton-type module on a heater block under vacuum. (b) Three-dimensional model. (c) Finite element mesh of the model. (d) $z$-position dependence of the temperature distribution when the hot side was set to \SI{393}{K}. ($x = \SI{0.0}{mm}$, $y = \SI{0.0}{mm}$). (e) Contour plot of the temperature distribution in the$x$-$y$ plane at $z = \SI{1.25}{mm}$.}
\label{fig6}
\end{figure*}

\red{The measured central values obtained using the HFM and GHM showed similar temperature-dependent trends and remained close over the investigated range up to a temperature difference of \SI{64}{K}. In contrast, the efficiency derived using the AC Harman method was systematically lower than those obtained using the HFM and GHM, and the discrepancy became more pronounced with increasing temperature difference.}

The close agreement between the values obtained using HFM and GHM is consistent with previous studies \cite{Rauscher2003,Rauscher2005}. The fitting curves based on Eq. (3) for the HFM and GHM results are shown as red and blue dotted lines, respectively, yielding corresponding $ZT$ values of 0.96 and 0.91. This comparison indicates that the $ZT = 0.81$ value derived from the AC Harman method is underestimated.

\subsection{Numerical simulation}
\red{To examine whether the observed discrepancy is consistent with differences in thermal boundary conditions, we calculated the steady-state temperature distribution under the open-boundary Harman configuration using COMSOL Multiphysics\cite{Buchalik2024,Yang2024}. Radiative heat exchange from the exposed module surfaces produces multidirectional heat-flow components and modifies the internal temperature distribution. Under the calculated conditions, the effective temperature difference across the thermoelectric region was smaller than the nominal temperature difference inferred from the heater temperature alone.} An overview of the model is shown in Fig. 6(a)-(c). The surface of a \SI{30}{mm} $\times$ \SI{30}{mm} Cu block ($z = \SI{0.0}{mm}$) was defined as the high-temperature heat source with $T_{\rm H} = \SI{393}{K}$. Figure 6(a) \red{shows} the $x$-$z$ plane of the 3D model. The surrounding environment was assumed to be a vacuum to eliminate convective heat transfer, and heat dissipation from the module occurred via thermal radiation in all directions. The finite element mesh consisted of approximately 135,000 nodes, as shown in Fig. 6(c).
 
\begin{figure}[t]
\centering
\includegraphics[width=7cm,clip]{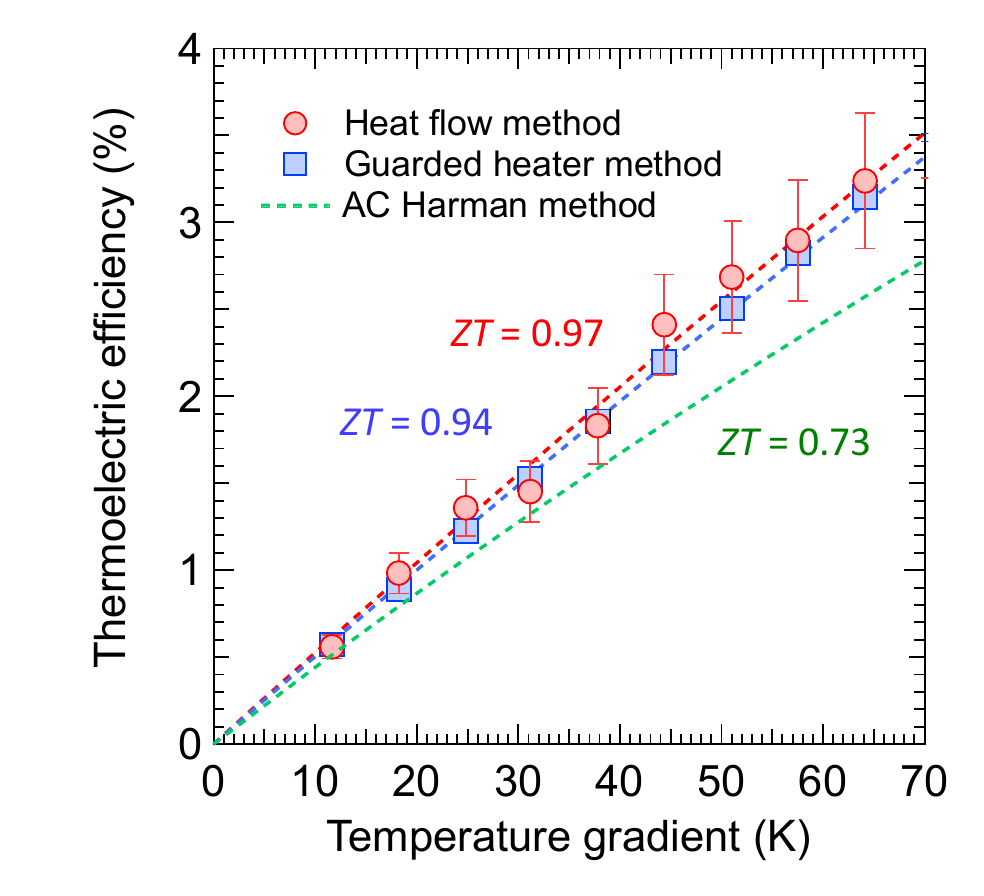}
\caption{Temperature different dependence of the thermoelectric conversion efficiency of the alumina-type module using the heat flow, guarded heater, and AC Harman methods. Red circles and blue squares denote the values obtained using the heat flow and guarded heater methods, respectively. The dotted lines represent the conversion-efficiency curves predicted by the model based on Eq. (3). The corresponding $ZT$ values were obtained from the fitting results using Eq. (3). \red{The HFM and GHM error bars indicate representative literature-based expanded uncertainty ranges adopted in this study, whereas the AC Harman error bars represent one sample standard deviation of five repeated measurements. Because these error bars have different definitions, they should not be interpreted as directly comparable confidence intervals. The AC Harman error bars are smaller than the symbol size (less than \SI{0.02}{\%}). The AC Harman values are efficiencies estimated from the measured effective $ZT$, whereas the HFM and GHM values were obtained from direct measurements of electrical power and heat input.}
}
\end{figure}

\begin{figure}[t]
\centering
\includegraphics[width=7cm,clip]{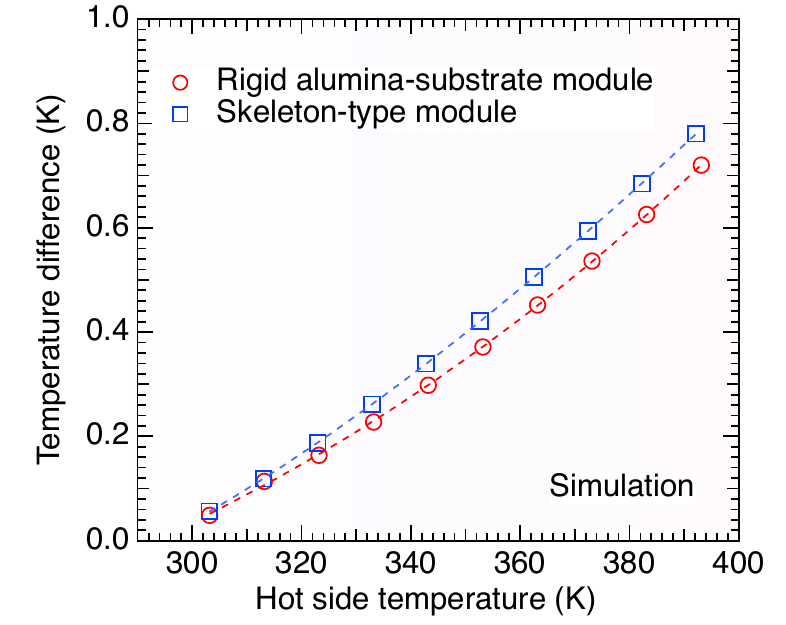}
\caption{\red{Calculated effective temperature difference across the thermoelectric material region under the open-boundary AC Harman configuration as a function of heater-block temperature. The calculated temperature difference represents the internal thermal state of the Harman measurement configuration and not a prescribed power-generation temperature difference.}}
\end{figure}
 
The calculated temperature distribution in the module \red{is} presented in Fig. 6(d) and (e). Figure 6(d) shows the temperature variation along the $z$ direction from the bottom of the module ($z = \SI{0.0}{mm}$) to the top ($z = \SI{1.35}{mm}$). Within the thermoelectric material region ($z = \SI{0.15}{mm}$ to \SI{1.25}{mm}), excluding the \SI{0.15}{mm}-thick polyimide substrate, the temperature varies linearly from \SI{391.05}{K} to \SI{391.80}{K}, corresponding to a temperature difference of \SI{0.75}{K}. Figure 6(e) shows the temperature distribution in the $x$-$y$ plane at $z = \SI{1.25}{mm}$, where a rapid temperature decrease is observed near the module edges, highlighting the effect of radiative heat loss. 

\red{The present FEM calculation represents the open thermal boundary used in the AC Harman configuration, rather than generator operation between prescribed hot-and cold-side temperatures. One surface of the module was maintained at the heater-block temperature, whereas the opposite surface exchanged heat only through the modeled parasitic pathways. Consequently, most of the module approached the heater-block temperature, and the effective temperature difference established across the thermoelectric material region was less than \SI{1}{K} under the representative condition. This value should not be interpreted as the temperature difference of a practical power generation efficiency measurement. Instead, it quantifies the limited internal temperature difference that develops in the open-boundary Harman configuration and demonstrates the difference between the nominal external temperature condition and the actual thermal state of the module.} 

\subsection{\red{Evaluation of modules with different substrate configurations}}
\red{
We next evaluated a commercial module with a rigid alumina substrate using the same procedure as that applied to the skeleton-type module. Figure 7 shows the temperature-different dependence of the conversion efficiency evaluated using the heat flow, guarded heater, and AC Harman methods. As observed for the skeleton-type module, the HFM and GHM results showed similar temperature-dependent trends and closely consistent central values. In contrast, the efficiency estimated from the AC Harman-derived effective $ZT$ was systematically lower than those obtained by the two direct efficiency measurements.}

\red{At $\Delta T = \SI{64}{K}$, the HFM and GHM central values differed by approximately  \SI{2}{\%} or less for both module configurations, whereas the efficiency derived using the AC Harman method was lower by approximately \SI{16}{\%} to \SI{30}{\%}. This discrepancy was substantially larger than the observed repeatability of the AC Harman measurements. Previous studies have demonstrated that Harman measurements are susceptible to nonideal contact and boundary conditions, electrical and thermal extrinsic effects, and parasitic heat transfer\cite{Castillo2010,Kang2016,Roh2016}. In particular, Roh et al. reported an underestimation of approximately \SI{10}{\%} to \SI{35}{\%} for thermoelectric modules measured under non-adiabatic conditions\cite{Roh2016}. The magnitude of the discrepancy observed in the present study is within this previously reported range, although the experimental conditions and underlying error mechanisms are not necessarily identical.
}

\red{The larger discrepancy observed for the rigid alumina-substrate module should not be attributed to the substrate material alone. The two commercial modules differ simultaneously in substrate configuration, substrate thickness, module dimensions, p-n pair number, electrode geometry, and exposed surface configuration. Therefore, the observed difference is more appropriately interpreted as a module-architecture-dependent effect rather than as an isolated material effect of alumina or polyimide.}

\red{To examine whether this interpretation is consistent with the thermal state expected under the AC Harman configuration, we compared the effective internal temperature differences obtained from the FEM calculations. Figure 8 shows the calculated temperature difference across the thermoelectric material region under the open-boundary Harman condition as a function of heater-block temperature. At representative heater temperatures of \SI{313.01}{K}, \SI{352.64}{K}, and \SI{392.13}{K}, the calculated effective temperature differences were \SI{0.12}{K}, \SI{0.42}{K}, and \SI{0.78}{K} for the skeleton-type module and \SI{0.11}{K}, \SI{0.37}{K}, and \SI{0.72}{K} for the rigid alumina-substrate module, respectively. These values are much smaller than the prescribed temperature differences used in the heat flow and guarded heater measurements, reflecting the fundamentally different thermal boundary condition of the AC Harman configuration.}

\red{The FEM results indicate that different module architectures establish different internal temperature distributions under identical open-boundary conditions, which is qualitatively consistent with the experimentally observed trend. However, the agreement should be regarded as qualitative support rather than a quantitative decomposition of the discrepancy into individual contributions such as substrate configuration, radiation, contact resistance, or geometry. The present steady-state FEM model does not explicitly include current-induced Peltier heating, Joule heating, transient thermoelectric response, or independently measured thermal contact resistance. Therefore, the FEM results are not used here to derive a universal correction factor for the AC Harman method.}

\red{The difference in the number of p-n pairs also does not, by itself, determine the ideal module figure of merit. For $N_{\rm pair}$ nominally identical p-n pairs connected electrically in series and thermally in parallel, the module Seebeck coefficient, electrical resistance, and thermal conductance scale as $N_{\rm pair}S_{\rm pair}$, $N_{\rm pair}R_{\rm pair}$, and $N_{\rm pair}K_{\rm pair}$, respectively. Consequently, the ideal module figure of merit is independent of $N_{\rm pair}$. In the present commercial modules, however, the p-n pair number covaries with leg geometry, electrode design, substrate structure, and exposed surface area. These factors cannot be separated using the present two-module comparison.}

\red{These results suggest that the AC Harman method is sensitive to architecture-dependent thermal boundary conditions when it is applied to commercial thermoelectric modules. In particular, radiative heat exchange, substrate-related heat spreading, contact resistance, and geometric effects can modify the effective internal temperature difference generated during the Harman measurement and thereby influence the apparent module-level $ZT$ and the efficiency estimated from it. For quantitative application of the AC Harman method to module-level efficiency evaluation, the effective internal temperature distribution should therefore be independently measured or evaluated using a module-specific thermal model. Nevertheless, because the method remains simple and highly reproducible, it is still useful for rapid performance screening, provided that the possible systematic bias associated with module architecture and boundary conditions is properly recognized.}

\red{In addition to radiative heat transfer, several factors including thermal and electrical contact resistance\cite{Sandoz2009,Bjork2015,Pablo2016,Xu2020,Ngan2015,Kanno2023,Li2024,Raju2024}, geometric effects\cite{Rowe1998,Ebling2010,Dunham2015,Ouyang2016,Zuo2023}, and temperature inhomogeneity\cite{Montecucco2014} may also contribute to the observed discrepancy and should be considered in future studies.}

\red{The present efficiency analysis was based on the conventional constant-property approximation. Incorporating the temperature dependence of thermoelectric properties may further improve the accuracy of the estimation, as discussed in recent studies\cite{Kim2015,Armstrong2016,Kim2017}. It should also be noted that the material figure of merit,  $zT = (S^{2}/\rho \kappa)T$, is not necessarily equivalent to the effective module-level $ZT$\cite{Snyder2017,Ryu2020,Ponnusamy2020}. Therefore, although the AC Harman method remains useful for rapid screening, quantitative module-level efficiency evaluation requires careful consideration of thermal boundary conditions and module architecture.}

\section{Conclusion}
A comparative study was conducted to evaluate three measurement methods for characterizing thermoelectric modules using commercial $\rm Bi_{2}Te_{3}$-based modules with different substrates. An experimental apparatus was developed to estimate the thermoelectric conversion efficiency using these methods. From a standardization viewpoint, our findings offer several important insights. 

\red{First, the heat flow method (HFM) and guarded heater method (GHM) yielded closely consistent central values and similar temperature-dependent trends over the investigated temperature range. Second, the efficiency estimated from the AC Harman measurement was systematically lower than the directly measured HFM and GHM efficiencies by approximately \SI{16}{\%} to \SI{30}{\%}. Finite-element analysis indicated that the open-boundary thermal condition used in the Harman configuration produces module-dependent internal temperature distributions that differ substantially from those established under controlled hot- and cold-side boundary conditions. These results highlight the importance of accounting for thermal boundary conditions and architecture-dependent heat-loss effects when applying the AC Harman method to module-level efficiency evaluation.}

\red{Importantly, the use of the same module specimens across the three measurement methods, together with systematic characterization of the boundary conditions and combined experimental and numerical analyses, provides a basis for practical measurement protocols, method-selection guidelines, and reference-module design, thereby supporting progress toward international standardization of thermoelectric module metrology.}

\section*{Acknowledgements}
This work was supported by a Grant-in-Aid for Research Activity Start-up (Grant No.17H07399) from the Japan Society for Promotion of Science (JSPS), Grant-in-Aid for Early-Career Scientists (Grant No.23K13552) from JSPS, the Thermal \& Electric Energy Technology Foundation (TEET), Iketani Science and Technology Foundation, and the Precise Measurement Technology Promotion Foundation (PMTP-F).

\section*{CRediT authorship contribution statement}
\textbf{Kenjiro Okawa:} Conceptualization, Data curation, Formal analysis, Funding acquisition, Investigation, Methodology, Visualization, Writing - original draft, Writing - review \& editing.
\textbf{Yasutaka Amagai:} Conceptualization, Data curation, Formal analysis, Funding acquisition, Investigation, Methodology, Project administration, Supervision, Visualization, Writing - review \& editing. 
\textbf{Norihiko Sakamoto:} Supervision, Writing - review \& editing. 
\textbf{Nobu-Hisa Kaneko:} Project administration, Supervision, Writing - review \& editing.

\section*{Data availability}
The data that support the findings of this study are available from the corresponding author upon reasonable request.


\bibliographystyle{elsarticle-num} 
\bibliography{TEMCOMP}
\end{document}